\documentclass[prl,aps,superscriptaddress,twocolumn,amssymb,showpacs]{revtex4-1}

\usepackage{color}
\usepackage{graphicx}
\usepackage{longtable}
\usepackage{epsfig}
\usepackage{dcolumn}
\usepackage{bm}
\usepackage{amssymb}

\def\beq{\begin{equation}}
\def\eeq{\end{equation}}
\def\bea{\begin{eqnarray}}
\def\eea{\end{eqnarray}}
\def\nn{\nonumber}

\begin{document}

\title{Frustration and Entanglement in Compass and Spin-Orbital Models}

\author {     Andrzej M. Ole\'{s} }
\email[Author's email: ]{a.m.oles@uj.edu.pl}
\affiliation{ Marian Smoluchowski Institute of Physics, Jagiellonian
              University, prof. S. \L{}ojasiewicza 11, PL-30348 Krak\'ow, Poland }
\affiliation{ Max-Planck-Institut f\"ur Festk\"orperforschung,
              Heisenbergstrasse 1, D-70569 Stuttgart, Germany }

\date{\today}

\begin{abstract}
We review the consequences of intrinsic frustration of the orbital 
superexchange and of spin-orbital entanglement. While Heisenberg 
perturbing interactions remove frustration in the compass model, 
the lowest columnar excitations are robust in the nanoscopic compass 
clusters and might be used for quantum computations. 
Entangled spin-orbital states determine the ground states in some 
cases, while in others concern excited states and lead to measurable 
consequences, as in the $R$VO$_3$ perovskites. 
On-site entanglement for strong spin-orbit coupling generates the 
frustrated Kitaev-Heisenberg model with a rich magnetic phase diagram 
on the honeycomb lattice. Frustration is here reflected in hole 
propagation which changes from coherent in an antiferromagnet via 
hidden quasiparticles in zigzag and stripe phases to entirely 
incoherent one in the Kitaev spin liquid. 
\end{abstract}


\maketitle

\subsection{1. Introduction}
\label{sec:intro}

Over the last decade the spin-orbital physics developed to a very 
active and challenging field which unifies frustrated magnetism 
and the phenomena in strongly correlated electron systems. It arose 
from the pioneering ideas of Kugel and Khomskii who recognized that 
spins and orbitals have to be treated on equal footing as quantum 
operators in transition metal oxides with partly filled degenerate 
$3d$ orbitals at large Coulomb interaction $U$ \cite{Kug82}. 
In Mott and charge-transfer insulators, the \textit{a priori} coupled 
spin-orbital degrees of freedom interact on the three-dimensional (3D) 
cubic lattice via the superexchange which follows from degenerate 
Hubbard model \cite{Ole83} and takes the form of a generalized 
Heisenberg model \cite{Ole05},
\begin{equation}
{\cal H}=\sum_{\langle ij \rangle\parallel\gamma}\!\left\{
  J^{(\gamma)}_{ij}(\vec\tau_i,\vec\tau_j) \vec{S}_i\!\cdot\!\vec{S}_j
+ K^{(\gamma)}_{ij}(\vec\tau_i,\vec\tau_j)\right\}.
\label{som}
\end{equation}
Here the operators $J^{(\gamma)}_{ij}$ and $K^{(\gamma)}_{ij}$ 
determine the Heisenberg exchange between spins 
${\vec S}_i\equiv\{S_i^x,S_i^y,S_i^z\}$ --- they depend on the bond 
direction $\gamma=a,b,c$ in the cubic lattice via the orbital operators 
$\{{\vec\tau}_i,{\vec\tau}_j\}$ at sites $i$ and $j$. Spin-orbital 
models (SOMs) relevant for real materials are quite involved and depend 
on whether the orbital degrees of freedom are $e_g$ or $t_{2g}$. 
They follow from virtual charge excitations along the bonds 
$\langle ij\rangle$ \cite{Ole05} and include the multiplet structure of 
excited states. Quantum fluctuations are of particular importance in 
$t_{2g}$ systems where two orbitals are active along each bond 
\cite{Kha05,Ole09}. In the case of large spins in the colossal 
magnetoresistance manganites with $S=2$ spins \cite{Dag01}, spins and 
orbitals nearly decouple and the $A$-type antiferromagnetic (AF) and 
ferromagnetic (FM) phase are well understood \cite{Fei99}. In spite of 
this decoupling of spins from orbitals, several questions remain, as for 
instance the theoretical explanation of the phase diagram of insulating 
manganites \cite{Goo06}. Even more challenging are systems with small 
spins, with their properties determined by spin-orbital entanglement 
(SOE) \cite{Ole12}.

While the intrinsic frustration of orbital interactions may be released 
by emerging spin-orbital order, the difference between spins and 
orbitals is best understood by considering generic orbital models, as 
the two-dimensional (2D) compass model \cite{vdB14} and the Kitaev 
model on the honeycomb lattice \cite{Kit06}. Both may be derived as 
limiting cases of magnetic interactions in Mott-Hubbard systems with 
partially filled $t_{2g}$ levels and with strong spin-orbit coupling 
\cite{Jac09} --- then SOE occurs on-site and 
leads to a rich variety of the low energy Hamiltonians that 
extrapolate from the Heisenberg to a quantum compass or Kitaev model. 
Yet, these two models are quite different --- the 2D compass model has 
one-dimensional (1D) nematic order at finite temperature \cite{Wen08}, 
while the exact solution of the Kitaev model is instead a disordered 
Kitaev spin liquid (KSL) with only nearest neighbor (NN) spin 
correlations. Realistic 2D or 3D $e_g$ orbital models are also strongly 
frustrated, but orbitals order at finite temperature following the 
strongest interactions \cite{Ryn10}, while quantum effects are small.

The purpose of this paper is to summarize selected recent developments 
presented at PM'14 Conference. We discuss the phase diagram of the 
compass-Heisenberg (CH) model in Sect. 2. Next we present a few 
examples of SOE in 1D and 2D systems, and in the $R$VO$_3$ perovskites 
(where $R$=Lu,Yb,$\dots$,La) in Sect. 3. 
The case of strong spin-orbit coupling realized in Na$_2$IrO$_3$ 
and frustrated interactions on the honeycomb lattice are analyzed 
in Sect. 4, The paper is summarized in Sect. 5.

\subsection{2. Frustration in Compass Models}
\label{sec:frust}

Although
the 2D Ising and compass model are in the same universality class, they
are quite different --- the first one is classical, while in the second 
one two pseudospin components $\{\tau_i^x,\tau_i^z\}$ interact either 
along horizontal or along vertical bonds by $J_x$ and $J_z$, 
and the ground state is highly degenerate and has 1D columnar order. 
Evolution between these two limits was investigated by the multiscale 
entanglement renormalization Ansatz and a quantum phase transition 
(QPT) from the 2D FM (AF) to nematic order was found close to the 
compass limit \cite{Cin10}. 

\begin{figure}[t!]
\includegraphics[width=8.2cm]{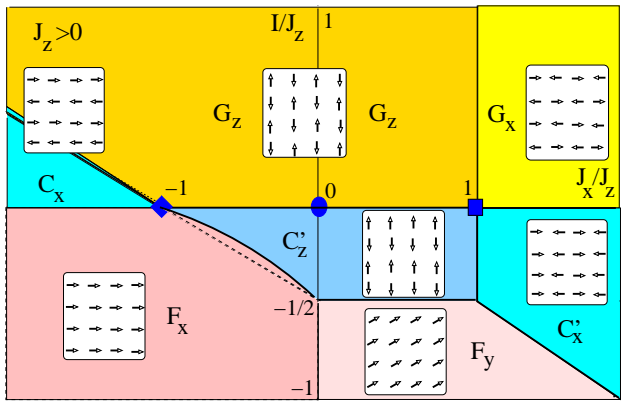}\\
\caption{
Phase diagram of the CH model in the $(J_x,I)$-plane for fixed AF
interaction $J_z=1$. Long-range spin order in phases 
$\{G_z,G_x,C'_z,C_x,C_x',F_x,F_y\}$ (the subscript $\alpha=x,y,z$ 
indicates the order parameter), depicted in a corresponding inset, 
replaces the nematic order for any finite $I$. Square ($J_x=J_z$) and 
diamond ($J_x=-J_z$) at the compass line ($I=0$) indicate multicritical 
points. The quantum corrections contribute to the QPTs between $F_x$ 
and $C'_z$ ($C_x$ and $G_z$) phases (solid lines).
This figure is reproduced from \cite{Tro10}.
}
\label{fig:phd}
\end{figure}

Another QPT occurs in the compass model itself for increasing 
$|J_x|/J_z$ at $|J_x|=J_z$, when the 1D order switches from vertical to 
horizontal bonds \cite{Oru09}. Understanding of symmetries in the 2D 
compass model allows one to calculate exact spectra of $L\times L$ 
clusters (with $L=6$) by mapping them to $(L-1)\times(L-1)$ clusters with 
modified interactions and to uncover the hidden dimer order \cite{Brz10}. 

The nematic order and the above hidden order in the 2D compass model 
are fragile and disappear in presence of infinitesimally small 
Heisenberg interaction $\propto I$ \cite{Tro10}. The CH
model (we take $J_z>0$),
\bea
H_{\rm CH}&=& J_x \sum_{i,j} \tau_{i,j}^{z} \tau_{i+1,j}^{z}
+ J_z \sum_{i,j} \tau_{i,j}^{x}\tau_{i,j+1}^{x} \nn\\
&+&I \sum_{i,j} \vec{\tau}_{i,j} \cdot
\left(\vec{\tau}_{i,j+1} + \vec{\tau}_{i+1,j}\right),
\label{CH}
\eea
has a very rich phase diagram (Fig. \ref{fig:phd}) and the symmetry 
breaking involves the component $\tau^{\alpha}_{i,j}$ with the 
strongest interactions. As both AF and FM interactions are possible, 
one finds also $C$-type AF ($C$-AF) order, with AF order between FM 
lines. The QPTs follow mostly from symmetry and are thus
given by straight lines. Surprisingly, however, the nematic order 
survives in the excited states in finite clusters, with somewhat 
lower quantum fluctuations for FM couplings $J_{\alpha}<0$. Indeed, 
this case should be of more importance for possible applications in 
quantum computing as information is easy to store by applying magnetic 
field when nematic order is FM. Crucial for these applications is 
large gap in spin excitations which occurs in the anisotropic $XYZ$ 
Heisenberg model (\ref{CH}). Therefore the columnar compass 
excited states are the lowest energy excitations in a broad range of 
parameters, when the perturbation $\propto I$ is weak and the cluster 
size is nanoscopic \cite{Tro10}. Certain realizations of computing 
devices with protected qubits have been implemented in Josephson 
junction arrays \cite{Gla09}, while systems of trapped ions in optical 
lattices look also promising \cite{Mil07}.

In the 1D compass model the consequences of frustration can be studied 
exactly, and one finds a QPT between two types of order on 
even/odd bonds at the $J_x=J_z$ point \cite{Brz07}. This model is 
quite distinct from the orbital $e_g$ model for a zigzag (ZZ) chain 
where frustration is weaker --- recent studies uncover rather peculiar 
behavior in the thermodynamic properties of the 1D compass model which 
follow from highly frustrated interactions \cite{You14}. An exact 
solution is also possible for a compass ladder \cite{Brz09}, which 
elucidates the nature of the QPT from ordered to disordered ground 
state found in the 2D compass model. Another type of frustration is 
encountered in the 1D plaquette compass model, where exact solution 
is no longer possible due to entanglement which increases locally 
in excited states and coincides with disorder \cite{BrzPC}. 

The 2D compass model can be seen as the strong-coupling limit of a 
spinless two-band Hubbard model with nonequivalent hopping matrices for 
the bonds along the $a$ and $b$ axis in the square lattice. Therefore,
a hole is not confined in the nematic state of the 2D compass model 
\cite{BrzDa}, unlike in the 2D Ising limit or in the 2D 
$t_{2g}$ orbital model \cite{Dag08}. The qualitative change of the hole 
excitation spectra near the nematic state corresponds to the QPT. 
An important common feature of the 2D orbital and compass model is that
quantum fluctuations are absent, and therefore the kinetic energy plays 
a particularly important role. It reorients the orbitals in the 2D 
alternating orbital (AO) state into ferro-orbital (FO) ordered domain 
walls that allow for deconfined motion of holes \cite{Wro10}, similar 
to FO order induced locally in a 1D doped $e_g$ system 
(manganite) \cite{Dag04}.

\subsection{3. Entanglement in Spin-Orbital Models}
\label{sec:enta}

Unless spins are FM, one has to consider orbitals coupled to spins in
the framework of general SOMs. In some cases the spin-orbital oder is 
determined by Kanamori-Goodenough rules stating the spin and orbital 
order are complementary, but in general SOE is expected. 
One of the main difficulties is a reliable approach to entangled ground 
states, as one can see on the example of frustrated exchange on the 
triangular lattice, where superexchange competes with direct exchange
\cite{Nor08}. In the disordered ground state with dimer orbital 
correlations SOE prevents any reliable predictions concerning the 
magnetic interactions on superexchange bonds, and spin correlations 
do not follow the sign of the spin exchange obtained using the 
mean-field (MF) approach \cite{Cha11}.

\begin{figure}[t!]
\begin{centering}
\includegraphics[clip,width=8.2cm]{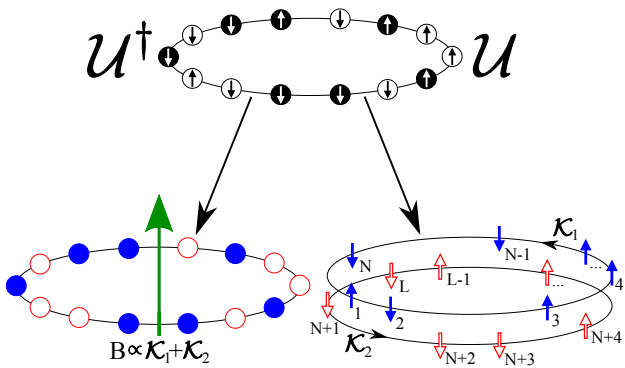} 
\par\end{centering}
\caption{  
Artist's view of the spin-orbital decoupling in the ring 
(\ref{ring}) caused by the transformation ${\cal U}$. 
The initial spin-orbital chain (top) splits into purely orbital 
(left) and spin (right) segments. The spin part consists of two 
halves carrying quasimomenta ${\cal K}_1$ and ${\cal K}_2$.
The orbital part feels an external magnetic field $\vec{B}$ 
perpendicular to the ring (arrow).
This figure is reproduced from \cite{Brz14}.
}
\label{fig:dec} 
\end{figure}

The SOE was discovered in 1D $d^1$ and $d^2$ systems with $t_{2g}$ 
orbitals \cite{Ole06}, but occurs also in $e_g$ systems, see below. The 
Bethe-Ansatz solution of the SU(4) 1D model \cite{Li99} demonstrates 
that its ground state and excitations are controlled by SOE. Recently 
another 1D model has been solved exactly providing a beautiful example 
of SOE, the SU(2)$\otimes XY$ ring \cite{Brz14},
\begin{equation}
{\cal H}_{{\rm SU(2)}\otimes XY}\!=\frac12 J 
\sum_{i=1}^{L}\big(\vec{\sigma}_i\cdot\vec{\sigma}_{i+1}\!+1\big) 
\big(\tau_{i+1}^{+}\tau_i^-\!+\tau_{i+1}^-\tau_i^{+}\big),
\label{ring}
\end{equation}
where $\sigma_l$'s are spin Pauli matrices, and $\tau_l$'s are orbital 
Pauli matrices, and $L+1\equiv 1$. The spin transposition operator, 
$X_{i,i+1}\equiv(\vec{\sigma}_i\cdot\vec{\sigma}_{i+1}\!+1)/2$,
interchanges spins on the bond $\langle i,i+1\rangle$, 
i.e., $X_{i,i+1}\vec{\sigma}_iX_{i,i+1}\!=\!\vec{\sigma}_{i+1}$.
For an open chain the spins and orbitals are decoupled by a unitary 
transformation ${\cal U}$ \cite{Kum13}, spins are disordered and the 
ground state has a large degeneracy of ${\cal D}=2^L$. Closing the 
spin-orbital chain to a ring (\ref{ring}) causes surprising changes in 
the spin part of the lowest-lying eigenstates. All the eigenstates are 
grouped in multiplets labeled by quasimomenta ${\cal K}$, 
and the ground state has ${\cal K}=0$ (Fig. \ref{fig:dec}). 
Therefore the topological order emerges and the 
ground state degeneracy drops to ${\cal D}=2^{L+1}/L$ \cite{Brz14}. 

When the orbital interactions have SU(2) symmetry,
\begin{equation}
{\cal H}_{{\rm SU(2)}\otimes{\rm SU(2)}}=\frac12 
J\sum_{\langle ij\rangle}
\big(\vec{S}_i\cdot\vec{S}_{i+1}+x\big)
\big(\vec{\tau}_i\cdot\vec{\tau}_{i+1}+y\big),
\label{SU(2)}
\end{equation}
one considers instead a 1D SOM (\ref{SU(2)}) with a higher SU(4) 
symmetry at $x=y=1/4$. Recently its phase diagram was investigated 
numerically for $J<0$ \cite{You12}. One finds four phases, with: FM/FO, 
AF/FO, AF/AO, and FM/AO order. The FM/FO ground state is disentangled, 
but SOE occurs in excited states. Spin and orbital excitations are 
entangled in the continuum, as well as a spin-orbital quasiparticle 
(QP) and bound state. A useful tool to investigate SOE in all these 
states is von Neumann entropy spectral function which gives the highest 
entanglement for the latter composite spin-orbital excitations, 
the QP and the bound state \cite{You12}. The scaling of the 
von Neumann entropy with system size is logarithmic and qualitatively 
different from other spin-orbital excitations from the continuum, where 
the entropy saturates.

Another example of SOE is found in the Kugel-Khomskii (KK) SOM, where
exotic types of magnetic order occur \cite{Brz12,Brz13}. The 2D KK 
model describes the superexchange $\propto J=4t^2/U$ in K$_2$CuF$_4$ 
between holes with $S=1/2$ spins in $e_g$ orbitals ($\tau=1/2$), 
\begin{eqnarray} 
\label{kk}
{\cal H}_{\rm KK}&=&\frac12 J\sum_{\langle ij\rangle||\gamma=ab}
\left\{-\,r_1\left({\vec S}_i\cdot{\vec S}_j+\frac{3}{4}\right)
\left(\frac{1}{4}-\tau^{\gamma}_i\tau^{\gamma}_j\right)\right. \nonumber \\
&+&\left. r_2\left({\vec S}_i\cdot{\vec S}_j-\frac{1}{4}\right)
\left(\frac{1}{4}-\tau^{\gamma}_i\tau^{\gamma}_j\right)\right.           \\
&+&\left. (r_2+r_4)\left({\vec S}_i\cdot{\vec S}_j-\frac{1}{4}\right)
\left(\frac{1}{2}-\tau^{\gamma}_i\right)
\left(\frac{1}{2}-\tau^{\gamma}_j\right)\right\}.              \nonumber
\end{eqnarray}
where $\tau^c_i=\tau^z_i=\sigma^z_i/2$, 
$\tau^{a,b}_i=(-\tau^z_i\pm\sqrt{3}\tau^x_i)/4$, while 
$r_1=1/(1-3\eta)$, $r_2=1/(1-\eta)$, and $r_4=1/(1+\eta)$ follow 
from the multiplet structure and depend on Hund's exchange 
$\eta\equiv J_H/U$. The second parameter extending the model (\ref{kk}) 
is the orbital splitting, $H_z=E_z\sum_{i}\tau_i^z$.

\begin{figure}[t!]
\includegraphics[width=8.2cm]{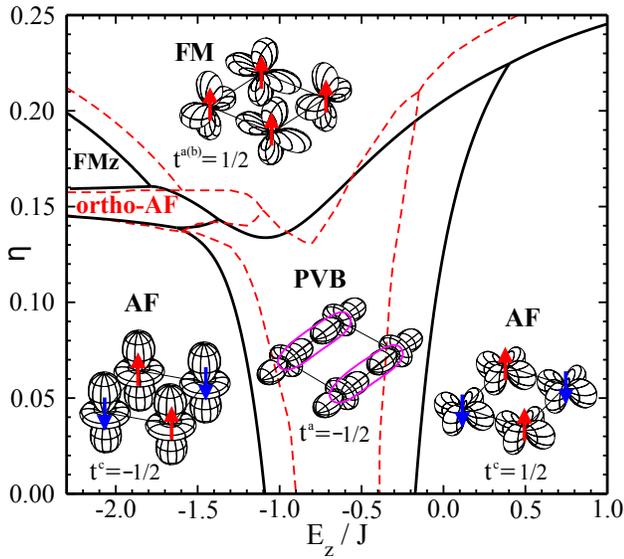}
\caption{ 
Phase diagram of the 2D KK model in the CMF and ERA
(solid and dashed lines).
Insets indicate spin-orbital configurations on a $2\times2$ plaquette 
--- $x$-like ($\tau^c_i\!=\!-1/2$) and $z$-like ($\tau^{c,a}_i\!=\!1/2$) 
orbitals are accompanied either by AF spin order (arrows) or by spin 
singlets (in the PVB phase). The FM phase has either a two-sublattice 
AO order or FO$z$ order (FM$z$). An exotic ortho-AF phase separates the 
AF and FM phases.
This figure is reproduced from \cite{Brz12}.
}
\label{fig:2d} 
\end{figure}

The phase diagram of the 2D KK model (\ref{kk}) obtained by two 
variational methods, a cluster MF (CMF) and entanglement renormalization 
Anzatz (ERA), contains the exotic magnetic order (ortho-AF phase in 
Fig. \ref{fig:2d}) between the AF and FM phase for $E_z<0$ and 
$\eta\simeq 0.155$ \cite{Brz12}, where the NN spin exchange changes 
sign. As shown in the perturbation theory which starts with $3z^2-r^2$ 
orbitals occupied by holes in the ground state and treats 
${\cal H}_{\rm KK}$ as perturbation, the next nearest neighbor (NNN) 
and third nearest neighbor (3NN) spin exchange is necessary to 
understand the origin of the four-sublattice AF phase 
(Fig. \ref{fig:2d}). This ground state is stabilized by local entangled 
spin-orbital excitations to spin singlets and $x^2-y^2$ orbitals 
\cite{Brz12}. Exotic magnetic order is also found in a bilayer and in 
the 3D KK model --- it follows again from SOE \cite{Brz13}.

In the $R$VO$_3$ perovskites SOE in excited states decides about the 
properties observed at finite temperature. While the spins and orbitals 
and their energy scales are well separated in the $R$MnO$_3$ 
perovskites \cite{Goo06}, the structural (orbital) and magnetic 
transition are here at rather similar temperature in $R$VO$_3$ 
\cite{Fuj10}. The orbital transition temperature $T_{\rm OO}=143$ K is 
almost the same as the N\'eel temperature $T_{\rm N1}=141$ K in 
LaVO$_3$, and increases with decreasing ionic radius $r_R$. Next it 
saturates and decreases from YVO$_3$ to LuVO$_3$ whereas $T_{\rm N1}$ 
decreases monotonically along the $R$VO$_3$ series. A theoretical 
explanation of these phase transitions requires the full superexchange 
model (\ref{som}) given in \cite{Kha01}, supplemented by the 
orbital-lattice interactions \cite{Hor08},
\begin{equation}
{\cal H}_{\rm orb}=E_z\sum_ie^{i{\vec R}_i{\vec Q}}\tau_i^z
+V_{ab}\sum_{\langle ij\rangle\parallel ab}\tau_i^z\tau_j^z
+g_{\rm eff}\sum_i\tau_i^x,
\label{rvo3}
\end{equation}
The leading term in the superexchange along the $c$ axis is similar to 
Eq. (\ref{SU(2)}), with spin $S=1$, $x=1$, $y=\frac14$, and orbital 
${\vec\tau}_i\equiv\{\tau_i^x,\tau_i^y,\tau_i^z\}$ operators 
($\tau=1/2$) for the active $t_{2g}$ orbitals, $yz$ and $zx$. 
The SOE occurs only along the bonds 
$\langle ij\rangle\parallel c$ as $xy$ orbitals are occupied at each 
site and thus orbital fluctuations are blocked along the bonds in the 
$ab$ planes. The crystal-field splitting term $\propto E_z$ supports 
$C$-type OO and alternates in the $ab$ planes, with 
${\vec Q}=(\pi,\pi,0)$. Actually, it competes with the superexchange 
which induces instead the observed $G$-type OO \cite{Fuj10}. 
The Jahn-Teller term $\propto V_{ab}$ supports as well AO order in the 
$ab$ planes, while along the $c$ axis FO order is favored by a similar 
interaction \cite{Hor08}, neglected for simplicity in Eq. (\ref{rvo3}). 

\begin{figure}[t!]
\includegraphics[width=8cm]{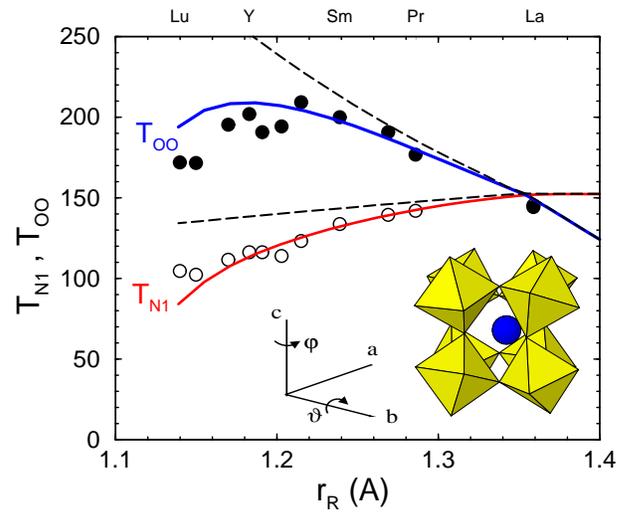}
\caption{ 
The orbital $T_{\rm OO}$ and N\'eel $T_{N1}$ transition temperature
(solid lines) for varying ionic size in $R$VO$_3$, as obtained from
the theory, and experimental points \cite{Fuj10} (full and empty 
circles). Dashed lines indicate $T_{\rm OO}$ and $T_{N1}$ 
obtained under neglect of orbital-lattice coupling ($g_{\rm eff}=0$). 
The inset shows the GdFeO$_3$-type distortion, with the rotation 
angles $\vartheta$ and $\varphi$ as in YVO$_3$. 
This figure is reproduced from \cite{Hor08}.
}
\label{fig:rvo}
\end{figure}

Due to SOE, which is activated in the excited states at finite 
temperature, it is crucial to employ a CMF approach, similar to the 
one used in KK models \cite{Brz13}. In this approach one determines
self-consistently the MF order parameters $\{\langle S_i^z\rangle$, 
$\langle\tau_i^z\rangle, \langle\tau_i^x\rangle,  
\langle S_i^z\tau_i^z\rangle\}$ by coupling a cluster along the $c$ 
axis to its neighbors via the MF terms adjusted to the 
$C$-AF/$G$-AO structure. The orbital fluctuations along the $c$ axis 
are very important and reduce significantly the orbital order 
parameter $\langle\tau_i^z\rangle$.

The structural transition at $T_{\rm OO}$ is explained as follows. In 
LaVO$_3$ the orthorhombic distortion $u\equiv(b-a)/a$ is small, where 
$a$ and $b$ are the lattice parameters of the $Pbnm$ structure. Then 
the values of $T_{\rm N1}$ and $T_{\rm OO}$ are used to establish the 
variation of model parameters with increasing lattice distortion $u$. 
All the parameters in ${\cal H}_{\rm orb}$ increase with increasing 
$u$ due to decreasing ionic size in $R$VO$_3$ (Fig. \ref{fig:rvo}). 
This increase is much faster for $g_{\rm eff}$ than for $E_z$ and 
$V_{ab}$, so from LaVO$_3$ to SmVO$_3$ the latter two parameters alone 
determine the increase of $T_{\rm OO}$. When $g_{\rm eff}$ becomes 
larger, however, this term acting as a field on the orbitals suppresses 
partly orbital order, $\langle\tau_i^z\rangle$, and the orbital 
polarization, $\langle\tau_i^x\rangle$, increases following local 
distortions. This reduces $T_{\rm OO}$ from YVO$_3$ to LuVO$_3$. At 
the same time $T_{\rm N1}$ decreases due to the changes in the orbital 
order. This decrease would not occur in the absence of lattice 
distortion (at $u=0$ implying $g_{\rm eff}=0$) which manifests again 
strong SOE in this system (Fig. \ref{fig:rvo}). We conclude that 
the lattice distortion $u$, which increases from La to Y by one order 
of magnitude, modifies orbital fluctuations and in this way tunes the 
onset of both orbital and spin order in the cubic vanadates.

There are more experiments which indicate strong SOE in the vanadium 
perovskites at finite temperature \cite{Ole12}. Here we mention briefly 
only the dimerization observed in the magnon spectra of the intermediate 
temperature $C$-AF phase in YVO$_3$ \cite{Ulr03}. Spin exchange 
interactions dimerize as a consequence of the instability of the 1D 
orbital chain along the $c$ axis. Of course, this mechanism cannot 
operate at $T=0$ as then the spins have rigid FM order along the $c$ 
axis. But thermal fluctuations in the spin system weaken spin 
correlations and dimerization is the way to lower the free energy. 
We emphasize that the dimerization occurs here simultaneously in both 
channels but the dimerization in the FM chain (for spins $S=1$) 
is much stronger than in the AO chain \cite{Sir08}. 

Summarizing, the SOE in the excited states is visible in the magnetic 
and optical properties of the vanadium perovskites, and any theoretical 
treatment has to go beyond a simple picture established by the 
Goodeough-Kanamori rules, and one has to go beyond this paradigm also 
in alkali $R$O$_2$ hyperoxides (with $R$=K,Rb,Cs) \cite{Woh11} or in 
finite clusters \cite{Bog10}. In the case of quantum states, these 
rules have to be generalized as follows: 
In the wave functions a component with spin-singlet and orbital-triplet 
coexists with a component with spin-triplet and orbital-singlet. 
SOE is also of importance for the pairing mechanism in Fe-pnictides 
\cite{Nic11}. More examples of SOE are presented in \cite{Ole12}.

\subsection{4. Strong spin-orbit coupling}
\label{sec:soc}

In systems with strong spin-orbit coupling, as in iridates, on-site SOE 
dominates and entangles locally spins and orbitals. In this case one 
has to determine first effective spins with eigenstates being linear 
combinations of spin-orbital components \cite{Jac09}. The interactions 
at low-energy between such effective $S=1/2$ spins are in general  
quite different from superexchange in spin-orbital models. Projecting 
the microscopic interactions on Kramers doublets gives strongly 
frustrated interactions \cite{Jac09}: 
($i$) the 2D compass model on the square lattice, and
($ii$) the Kitaev model on the honeycomb lattice. Kitaev model is a 
realization of a spin liquid with only NN spin correlations on a 
nonfrustrated lattice \cite{Kit06}. Similar to the Kitaev model, the 
triangular lattice of magnetic ions in an ABO$_2$ structure 
(as for instance in LiNiO$_2$ \cite{Rei05}) has Ising-like interactions 
$\propto S_i^{\alpha}S_j^{\alpha}$ with $\alpha=x,y,z$ for three 
nonequivalent bond directions in the lattice. In what follows we focus 
on the magnetic interactions on the honeycomb lattice which attracted 
a lot of attention recently. 

Here we consider the Kitaev-Heisenberg (KH) $t$-$J$ model ($J>0$) on 
the honeycomb lattice with two sublattices $A$ and $B$ \cite{Bas07}, 
realized in Na$_2$IrO$_3$ \cite{Cho12,Com12},
\begin{eqnarray}
&{\cal H}_{tJ}&\,\equiv
t\sum_{\langle ij\rangle\sigma} c^{\dagger}_{i\sigma}c^{}_{j\sigma}
 + J_K\sum_{\langle ij\rangle\parallel\gamma} S_i^\gamma S_j^\gamma
 + J_1\sum_{\langle ij\rangle}   \vec{S}_i\!\cdot\!\vec{S}_j \nonumber \\
&+&\!(1-\alpha)\,\Big\{
   J_2\!\!\sum_{\{ij\}\in{\rm NNN}}\!\!\vec{S}_i\!\cdot\!\vec{S}_j
 + J_3\!\!\sum_{\{ij\}\in{\rm 3NN}}\!\!\vec{S}_i\cdot\vec{S}_j\Big\}\,,
\label{KH}
\end{eqnarray}
with FM Kitaev and AF Heisenberg NN exchange,
\begin{equation}
J_K\equiv -2J\alpha\,, \hskip .7cm J_1\equiv J(1-\alpha)\,.
\label{JJ}
\end{equation}
The parameter $0\le\alpha\le 1$ interpolates between the Heisenberg and 
Kitaev model. The NN Kitaev ($J_K$) and Heisenberg ($J_1$) interactions 
compete and the spin order changes with increasing $\alpha$. 
The signs of these two competing terms (\ref{JJ}) are opposite and both 
AF/FM and FM/AF Heisenberg/Kitaev were studied \cite{Cha10}. Such spin 
interactions were proposed to describe the Mott-insulating layered 
iridates \cite{Kim11} --- for $J_1>0$ also NNN ($J_2$) and 3NN ($J_3$) 
Heisenberg terms are necessary as only then the experimentally observed 
ZZ magnetic order in Na$_2$IrO$_3$ \cite{Cho12} is reproduced. The term 
$\propto t$ stands for the kinetic energy of composite fermions with 
pseudospin flavor $\sigma$ in the restricted space which contains no 
double occupancies. 

\begin{figure}[t!]
\begin{center}
\includegraphics[width=6.2cm]{s_J04.eps}  \vskip -1.0cm
\includegraphics[width=7.0cm]{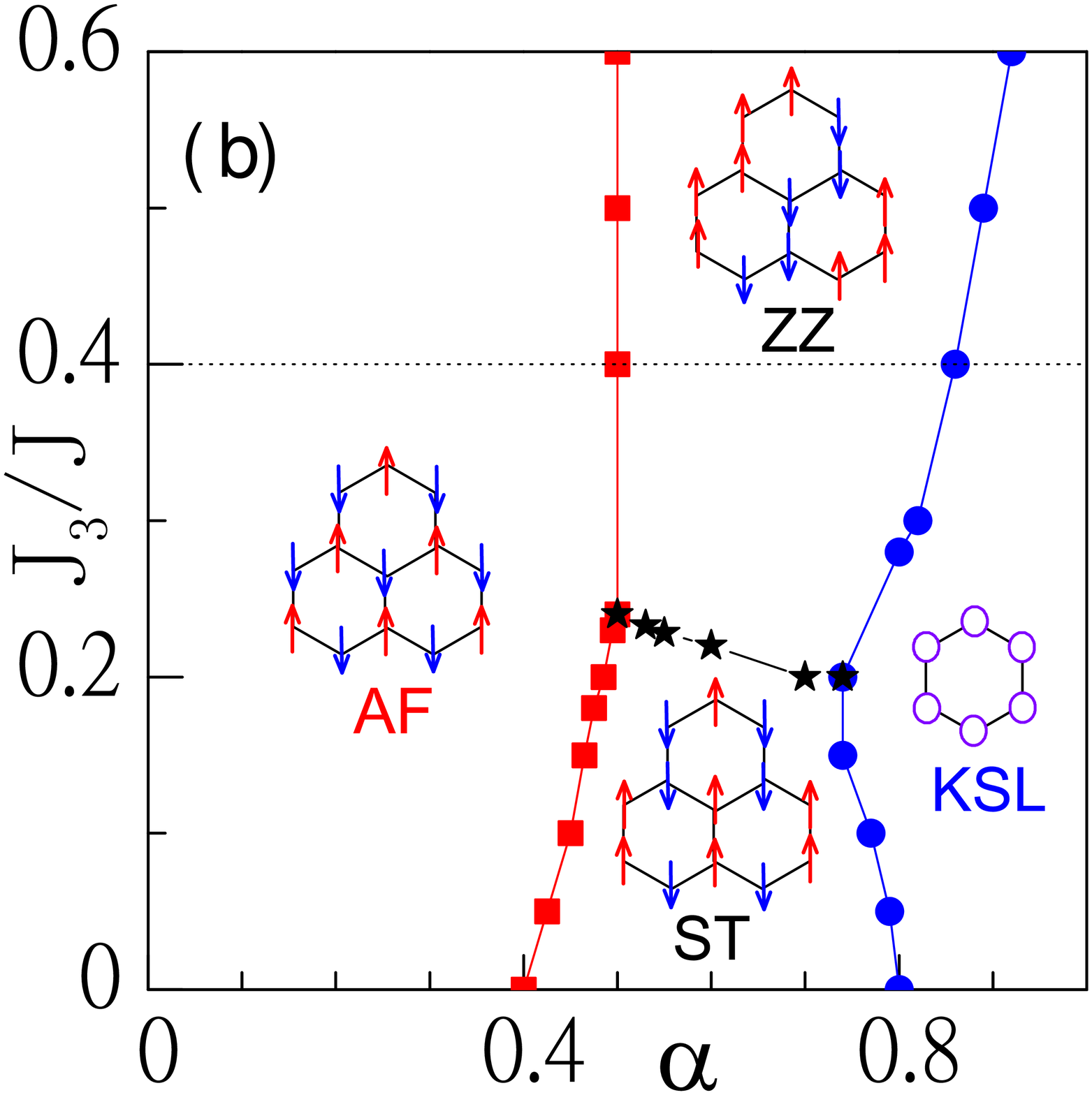}\\
\end{center}
\vskip -.2cm
\caption{ 
Magnetic phases in the KH model Eq. (\ref{KH}) shown in: 
(a) spin correlations ${\cal S}_{\Phi}$ (\ref{Phi}) representing the AF 
and ZZ order ($\Phi$=AF, $\Phi$=ZZ), and the Kitaev invariant ${\cal L}$ 
(\ref{L}) for $J_2=0$, $J_3=0.4J$;
(b)~phase diagram in the $(\alpha,J_3/J)$ plane (points) for $J_2=0$, 
with AF, ST, ZZ spin order and KSL disordered phase. 
The insets show spin order (arrows) or disorder (circles).  
This figure is reproduced from \cite{Tro14}.
}
\label{fig:sl}
\end{figure}

Consider first a spin order parameter for a phase $\Phi$,\\
\begin{equation}
{\cal S}_{\Phi}^2\equiv\frac{12}{N^2}\sum_{ij}
e^{i\vec{k}\cdot(\vec{R}_i-\vec{R}_j)}
\langle(S^z_{iA}\pm S^z_{iB})(S^z_{jA}\pm S^z_{jB})\rangle),
\label{Phi}
\end{equation}
where the average is calculated in the ground state $|\Phi\rangle$. 
Investigating ${\cal S}_{\Phi}$ allows one to identify the symmetry 
breaking and long-range spin order studying finite clusters where the 
symmetry broken states do not occur \cite{Kap89}. In the above 
definition the signs of the spin components $S_{jB}^z$ on sublattice 
$B$ and the vector $\vec{k}$ are selected differently, depending on the 
spin order in the considered magnetic phase $\Phi$ \cite{Tro14}. 
Investigating such spin correlations does not suffice to identify the 
disordered KSL, with finite NN spin correlations. Here we evaluate 
instead the Kitaev invariant \cite{Kit06} for a single 
hexagon ${\cal C}_6$, 
\begin{equation}
{\cal L}\equiv 2^6\left\langle\prod_{i\in{\cal C}_6}S^\gamma_i\right\rangle.
\label{L}
\end{equation}
At fixed $J_3=0.4J$ one finds first large AF spin correlations  
${\cal S}_{\rm AF}$ for $\alpha<0.5$, and next the ZZ phase is favored 
for $\alpha>0.5$, as indicated by large ${\cal S}_{\rm AF}$ or 
${\cal S}_{\rm ZZ}$ [Fig. \ref{fig:sl}(a)]. The AF$\leftrightarrow$ZZ 
transition at $\alpha=0.5$ follows from symmetry and is independent of 
the cluster size. Both ${\cal S}_{\rm AF}$ and ${\cal S}_{\rm ZZ}$ 
decrease for $\alpha>0.85$ when the ground state of the KH model 
(\ref{KH}) approaches the KSL at $\alpha\to 1$, and ${\cal L}\to 1$. 

The phase diagram of the KH model in the $(\alpha,J_3/J)$ plane was 
determined \cite{Tro14} by analyzing spin order parameters, 
${\cal S}_{\Phi}^2$, and the fidelity susceptibility. It contains 
four magnetic phases at $J_2=0$, the AF, ZZ, stripe (ST) phase, 
and the KSL disordered phase [Fig. \ref{fig:sl}(b)]. The AF phase is 
stable for small $\alpha$, while at intermediate $\alpha$ it is 
replaced by two other magnetic phases, ST and ZZ.. These types of 
order with coexisting AF and FM bonds manifest enhanced frustration 
for increasing Kitaev interactions. The gapless KSL takes over at 
$\alpha>0.85$ and is also stable in presence of lattice distortions 
\cite{Sel14}. As shown in the CMF approach, the AF phase is also 
destabilized by increasing NNN interactions $J_2$ \cite{Alb11}. 
The phase diagram of the KH model (\ref{KH}) is investigated by 
several groups at present; further results are given in \cite{Got14}.

Motivated by strongly incoherent photoemission (PES) spectra found in 
Na$_2$IrO$_3$ \cite{Com12}, we have used Lanczos diagonalization of the 
$N=24$ site cluster with periodic bondary conditions to study the 
evolution of hole spectral functions for varying interactions in the KH
model (\ref{KH}) \cite{Tro14}. Indeed, the spectra observed in PES for 
Na$_2$IrO$_3$ are rather unexpected as in spite of ZZ spin order, 
no QPs and only incoherent spectra are observed. A systematic study of 
hole spectral properties in this magnetic phase requires to consider 
two distinct Green's and two spectral functions \cite{Tro13}:
($i$) full spectral function which corresponds to the PES, and
($ii$) the sublattice spectral function, i.e., when a hole moves over 
sites of one sublattice only. 
One finds that in the ZZ phase the QPs appear only in the sublattice 
spectral function while they are hidden in the PES spectral function. 
This result is independent of the model used to stabilize the ZZ order 
and may be considered as following from symmetry of the honeycomb 
lattice which supports destructive interference in the PES spectral 
function at low energy \cite{Tro13}. 

As expected, one finds coherent QPs in the PES spectral function for 
the AF phase at weakly frustrated interactions, but a similar 
interference and hidden QPs are found in the ST phase \cite{Tro14}.
In is interesting to ask what will happen when spin iteractions are 
maximally frustrated and the ground state is the KSL. A naive argument 
that in the absence of robust spin order, quantum spin fluctuations 
will not couple to the moving hole to generate coherent propagation 
turns out to be correct, and using exact diagonalization one finds 
indeed no coherent QPs here, also in the sublattice spectral function. 
This result was derived by analyzing the sublattice spectral function 
in absence of spectral broadening \cite{Tro14}, $A_d(\vec{k},\omega)=
\sum_n\alpha_d({\vec k},\omega_n)\delta(\omega-\omega_n)$.
The spectral weights $\{\alpha_d({\vec k},\omega_n)\}$ are totally 
incoherent at low excitation energies $\omega_n$, except at the 
$\vec{k}=\Gamma$ point in the momentum space \cite{Tro14}. This 
important result follows from the presence of vortex gap in the 
Majorana excitations \cite{Bas07} and implies that Ising-like NN spin 
correlations in the KSL phase are insufficient to generate coherent 
hole propagation and thus carrier motion in the lightly doped KSL is 
non-Fermi liquid like. This analysis allows one to conclude 
\cite{Tro14} that gapless Majorana excitations are responsible for the 
absence of QPs in the close vicinity of the $\Gamma$ point and, on the 
contrary to some earlier claims, the weakly doped KSL is \textit{not}  
a Fermi liquid.

\subsection{5. Summary and Outlook}
\label{sec:summa}

We have shown that orbital superexchange interactions have lower 
symmetry than spin ones --- they are directional and intrinsically 
frustrated, also on geometrically nonfrustrated lattices. This leads to 
nematic order and provides new opportunities for quantum computing. The 
nematic order in the 2D compass model is robust and survives in excited 
states which could be used for storing information in nanoscopic 
systems, where Heisenberg perturbing interactions remove frustration 
and trigger long-range order in the ground state.  

In spin-orbital systems frustration in the orbital channel is 
frequently removed by spin order which modifies the exchange in the 
orbital subsystem. Nevertheless, SOE is characteristic in these 
systems and may have measurable consequences at finite temperature,
as in the $R$VO$_3$ perovskites. Here, similar to 1D spin-orbital model 
systems, frustration and entanglement occur simultaneously. In contrast,
in systems with strong spin-orbit interaction entanglement comes first 
and generates frustration as shown on the example of the KH model on 
the honeycomb lattice. We have also shown that the spectral functions 
obtained from the KH model with frustrated interactions describe hidden 
QPs in the ordered phases with coexisting FM and AF bonds,
while the Ising-like short-range spin correlations in the KSL are 
insufficient to generate coherent hole propagation. 

We have presented only selected recent developments in the field of 
spin-orbital physics. Among others, we would like to mention models 
which describe interfaces or heterostructures, and hybrid bonds
between ions with different fillings of the $d$ shell. For instance, 
$d^3$ impurities generate also frustrated interactions in the 
$d^3$-$d^4$ system. Having no orbital degree of freedom, they are able 
to modify locally orbital order, and the actual spin-orbital order in 
ruthenates might even totally change at finite doping \cite{arXiv}. 
Summarizing, the spin-orbital physics is a very active and fast 
developing field of frustrated magnetism with numerous challenging 
and timely problems, both in the experiment and in the theory. 
We apologize for not including here many other interesting 
developments in this field due to the lack of space.

\subsection{Acknowledgments}

It is our pleasure to acknowledge collaboration and  
valuable discussions with 
W.~Brzezicki, J.~Chaloupka, L.~Cincio, M. Cuoco, M. Daghofer, 
J.~Dziarmaga, R.~Fr\'esard, A. Herzog, B. Normand, J.~Sirker, 
J. van den Brink, F. Trousselet, K. Wohlfeld, and W.-L. You. 
I thank particularly warmly 
Louis Felix Feiner, Peter Horsch, and Giniyat Khaliullin
for a very friendly and insightful collaboration.
We kindly acknowledge financial support by the Polish National 
Science Center (NCN) under Project No. 2012/04/A/ST3/00331.



\end{document}